# TOMOGRAPHIC PROBABILITY REPRESENTATION FOR STATES OF CHARGE MOVING IN VARYING FIELD

V.I. Man'ko[1,2], E.D. Zhebrak[1]

[1] *Moscow Institute of Physics and Technology*

[2] *Lebedev Physical Institute of the Russian Academy of Sciences*

## Abstract

The coherent and Fock states of a charge moving in varying homogeneous magnetic field are studied in the tomographic probability representation of quantum mechanics. The states are expressed in terms of quantum tomograms. The coherent states tomograms are shown to be described by normal distributions with varying dispersions and means. The Fock state tomograms are given in the form of probability distributions described by multivariable Hermite polynomials with time-dependent arguments.

## Introduction

The classical-like packets [1] for charge moving in constant homogeneous magnetic field were shown [2] to be the coherent states with constant dispersions and means moving along the classical cyclotron trajectory and the coherent state wave functions were also constructed for the charge moving in varying uniform magnetic field [3], [4], [5]. In [6] the coherent states were used to obtain the Landau diamagnetism by new method.

The probability representation of quantum mechanics was suggested recently [7] for the case of continuous variables like position and momentum, and in [8], [9] for discrete variables like spin. Tomographic probability distributions for position and spin variables are related to wave function by means of invertible integral transforms and can be used to describe quantum states for all systems.

Generalized coherent states in [10] were introduced to describe the quasiclassical motion of electrons in a microwave and in a homogeneous magnetic field. In [11] the coherent states were considered for the quantum particle on a circle. The obtained results were applied in [12] to a problem of a charged particle moving in a uniform magnetic field. A special type of stationary states for a particle in a homogeneous magnetic field – so-called squeezed states were considered for example in [13] and [14]. In [15] were introduced so-called trajectory coherent states corresponding to Gaussian wave packets with center of gravity moving along the classical trajectories of charged particles. Complete sets of trajectory coherent states were derived for Schrodinger, Klein-Gordon and Dirac equations in [16] – [18]. In the purpose of study the rotational motion of a self-bound system of particles, such as a nucleus coherent states of the rotation groups are constructed by several authors (see [19]). The coherent states construction algorithms were proposed in [2] (for systems with quadratic Hamiltonians with discrete spectra), in [20] where the approach introduced in [2] was generalized to systems with continuous spectra and in [21] (considering a charge in electromagnetic field).

The problem of charges moving in magnetic field has different applications. In [22], [23] the tunneling in quantum dots is considered in terms of Kondo effect. In [24] formation of a coherent Kondo – singlet state and stabilization of a spin liquid in a Kondo lattice is considered. In [25] were introduced two-scale magnetic Wannier orbitals of electrons in a von Neumann lattice. Transport properties of two-dimensional electron gases under high perpendicular magnetic fields were studied in [26]. In [27] and [28] correspondingly was introduced a description of charged electron-hole complexes in magnetic fields and considered a system of two electrons and one hole in a strong magnetic field, i.e. a negatively charged magnetoexciton. In the paper [29] was studied the ground Landau level of the 2D Frohlich polaron in a magnetic field and in [30] the polarization properties of an atomic gas in a coherent state were presented.

In our previous work [31] the coherent states of the charge moving in constant magnetic field were studied in the tomographic probability representation. The aim of this work is to extend the consideration of probability representation of the problem and to find tomographic distributions (tomograms) describing the quantum states of charged particle moving in magnetic field to the case of time-varying fields. Our goal is to obtain the tomograms for both coherent states and Fock states of the charged particle in explicit form.

We remind the discussion of quantum state concept which is the basic concept of quantum theory. In conventional formulation of quantum mechanics the notion of system state is associated with wave function $\psi(x)$ [32] or density matrix [33], [34]. These complex functions $\psi(x)$ or $\rho(x,x')$ differ essentially from the notion of a system state in classical statistical mechanics where the nonnegative probability density $f(q,p)$ on the system phase space is identified with the state concept. In view of this from the very beginning of quantum mechanics some attempts to get a formulation of quantum mechanics where the notion of system state is similar to classical probability density $f(q,p)$ were made [35], [36], [37], [38].

A solution of this problem was also suggested by Pauli [39] who proposed a conjecture that the quantum state complex wave function can be reconstructed if one knows both the probability density of position $|\psi(x)|^2$ and probability density of momentum $|\tilde{\psi}(p)|^2$. Though this conjecture was shown to be incorrect (see, e.g. [40]) the generalization of this suggestion which now is called probability representation of quantum mechanics or tomographic probability representation solved this problem [7]. One can find such probability distribution $w(X,\mu,\nu)$ of position $X$, depending on extra real parameters $\mu$ and $\nu$ that the density matrix or wave function can be obtained from this distribution by means of an integral transform. In our work we apply this new description of quantum state to the problem of charge moving in varying fields.

The paper is organized as follows. In Sec. 2 a short review of tomographic probability representation for systems with one and two degrees of freedom is presented.

In Sec.3 the construction of coherent states and its tomograms for charged particle moving in constant magnetic field by means of creation and annihilation operators is reminded ([3], [41]). In Sec. 4 the Gaussian states of quantum charge including coherent states are considered in tomographic probability representation of quantum mechanics. In Sec. 5 some conclusions and perspectives are presented.

*Quantum tomograms*

We construct the tomograms of the coherent states of a charge moving in varying fields and the tomograms identified with the quantum states determined by time-dependent integrals of motion which are initial energy and initial orbital momentum possessing the discrete time-independent eigenvalues.

The quantum tomography is based on the map of density operator $\hat{\rho}$ of the state $|\psi\rangle$ onto tomographic probability distribution $w(X,\mu,\nu)$ called symplectic tomogram determined by the relation:

$$w_s(X,\mu,\nu) = Tr\hat{\rho}\cdot\delta(\hat{X} - \mu\hat{q} - \nu\hat{p}). \qquad (1)$$

Here we consider a system with one degree of freedom and $X$, $\mu$, $\nu$ are reals, $\hat{q}$ and $\hat{p}$ are position and momentum operators respectively.

The tomogram is nonnegative probability distribution of random variable $X$ which is position in rotated and rescaled reference frame in the phase-space. The parameters of the reference frame are labeled by reals $\mu$ and $\nu$. The inverse of (1) reads

$$\hat{\rho} = \frac{1}{2\pi} \int w(X,\mu,\nu) e^{i(X-\mu\hat{q}-\nu\hat{p})} dX d\mu d\nu. \tag{2}$$

The tomogram has homogeneity property

$$w(\lambda X, \lambda\mu, \lambda\nu) = \frac{1}{|\lambda|} w(X,\mu,\nu). \tag{3}$$

For normalized states $\hat{\rho}$ the tomographic probability distribution is normalized i.e.

$$\int w(X,\mu,\nu) dX = 1. \tag{4}$$

In case of $\mu = \cos\theta$, $\nu = \sin\theta$ the tomogram

$$w_{op}(X,\theta) = w(X,\cos\theta,\sin\theta) \tag{5}$$

is called optical tomogram.

The symplectic tomogram of pure state with the wave function $\psi(y)$ is determined by the formula [42]:

$$w(X,\mu,\nu) = \frac{1}{2\pi|\nu|} \left| \int \psi(y) e^{\frac{i\mu}{2\nu}y^2 - \frac{iX}{\nu}y} dy \right|^2, \tag{7}$$

which is related to fractional Fourier transform of the wave function.

For the several degrees of freedom the formulas (1-7) are given as simple direct product of the transforms presented for one-degree freedom. For example for two degrees of freedom the symplectic tomogram $w(X_1,\mu_1,\nu_1,X_2,\mu_2,\nu_2)$ is determined by the fractional Fourier transform of the wave function $\psi(y_1,y_2)$ and it reads

$$w(X_1,\mu_1,\nu_1,X_2,\mu_2,\nu_2) = \frac{1}{4\pi^2|\nu_1\nu_2|} \left| \int \psi(y_1,y_2) e^{\frac{i\mu_1}{2\nu_1}y_1^2 + \frac{i\mu_2}{2\nu_2}y_2^2 - \frac{iX_1}{\nu_1}y_1 - \frac{iX_2}{\nu_2}y_2} dy_1 dy_2 \right|. \tag{8}$$

The optical tomogram $w_{op}(X_1,X_2,\theta_1,\theta_2)$ is given by Eq(8) where $\mu_1 = \cos\theta_1$, $\mu_2 = \cos\theta_2$, $\nu_1 = \sin\theta_1$, $\nu_2 = \sin\theta_2$. If one has two states $\hat{\rho}_1$ and $\hat{\rho}_2$ the fidelity providing transition probability $P_{12} = P_{21}$ between these states reads

$$P_{12} = Tr\hat{\rho}_1\hat{\rho}_2 = \frac{1}{4\pi^2} \int w_1(X_1,\mu_1,\nu_1,X_2,\mu_2,\nu_2) w_2(Y_1,\mu_1,\nu_1,Y_2,\mu_2,\nu_2) e^{i(X_1-Y_1+X_2-Y_2)} dX_1 dY_1 d\mu_1 d\nu_1 d\mu_2 d\nu_2 \tag{9}$$

We can use this formula to study the transition probabilities between the energy levels of the charge excited by time-varying field.

## Coherent states of charged particle moving in constant magnetic field

We shall consider a particle of mass $m=1$ and charge $e=1$ moving in a constant magnetic field with a vector potential

$$\vec{A} = \frac{1}{2}\left[\vec{H} \times \vec{r}\right]. \tag{10}$$

Let the cyclotron frequency $\omega = 1$. The Hamiltonian for such a system is

$$H = \frac{1}{2}\left[(p_x - A_x)^2 + (p_y - A_y)^2\right], \quad \hbar = c = e = 1. \tag{11}$$

In order to construct the coherent and excited states of the charge moving in the constant magnetic field let's introduce the following operators:

$$\hat{A} = \frac{(p_x + ip_y) + \frac{1}{2}(y - ix)}{\sqrt{2}}, \quad \hat{B} = \frac{(p_y + ip_x) + \frac{1}{2}(x - iy)}{\sqrt{2}}. \tag{12}$$

The following commutation relations hold:

$$\left[\hat{A}, \hat{A}^+\right] = \left[\hat{B}, \hat{B}^+\right] = 1, \quad \left[\hat{A}, \hat{B}\right] = \left[\hat{A}, \hat{B}^+\right] = 0 \tag{13}$$

For oppositely charged particles, the lowering and raising operators $\hat{A}, \hat{B}$ and $\hat{A}^+, \hat{B}^+$ change their places. For simplicity we suppose $e > 0$.

The Hamiltonian of the system can be expressed in terms of the introduced operators and it reads

$$H = \hat{A}^+\hat{A} + \frac{1}{2}. \tag{14}$$

This operator is apparently the integral of motion. Also the operator $\hat{A}e^{it}$ is the integral of motion.

Because of the axial symmetry of the electromagnetic field potential (10), the $z$ component of the angular momentum is also the integral of motion and it may be also expressed in terms of our operators (12):

$$L_z = \hat{B}^+\hat{B} - \hat{A}^+\hat{A}. \tag{15}$$

The operators (12) are related specific integrals of motion determining a center of cyclotron motion, namely:

$$\hat{A} = \frac{[y - y_0 - i(x - x_0)]}{\sqrt{2}}, \quad \hat{B} = \frac{[x_0 - iy_0]}{\sqrt{2}}, \tag{16}$$

where

$$x_0 = \frac{1}{2}x + p_y, \quad y_0 = \frac{1}{2}y - p_x$$

are the well-known preserving "coordinates" of the center of the orbit of a particle moving in a constant magnetic field.

It is known from [2] that the eigenvalue of the invariant $\hat{B}$ determines the coordinates of the center of the orbit in the $xy$ plane, and the eigenvalue of the operator $\hat{A}$ determines the current coordinates of the center of the wave packet.

The operators $\hat{A}$, $\hat{B}$ have the normalized eigenvectors called coherent states $|\alpha, \beta\rangle$ that obey the Schrodinger equation. The following formulas hold:

$$\hat{A}|\alpha, \beta\rangle = \alpha|\alpha, \beta\rangle, \quad \hat{B}|\alpha, \beta\rangle = \beta|\alpha, \beta\rangle. \tag{17}$$

Since we have two lowering operators $\hat{A}$ and $\hat{B}$ with bosonic commutation relation the coherent states will depend on two parameters:

$$|\alpha, \beta\rangle = \exp\left[-\frac{1}{2}\left(|\alpha|^2 + |\beta|^2\right)\right] \sum_{n_1, n_2 = 0}^{\infty} \frac{\alpha^{n_1} \beta^{n_2}}{\sqrt{n_1! n_2!}} |n_1, n_2\rangle. \tag{18}$$

where $\alpha, \beta$ are arbitrary complex numbers. Here

$$|n_1, n_2\rangle = \frac{(A^+)^{n_1} (B^+)^{n_2}}{\sqrt{n_1! n_2!}} |0, 0\rangle$$

and these state are the Fock states of the charge moving in the magnetic field. The ground state $|0, 0\rangle$ satisfying the condition

$$A|0, 0\rangle = B|0, 0\rangle = 0$$

is solution of the Schrodinger equation $H|0, 0\rangle = \frac{1}{2}|0, 0\rangle$. At the same time the states $|n_1, n_2\rangle$ are eigenstates of the Hamiltonian $H$ and the angular momentum $L_z$:

$$H|n_1, n_2\rangle = \left(n_1 + \frac{1}{2}\right)|n_1, n_2\rangle, \tag{19}$$

$$L_z|n_1, n_2\rangle = (n_2 - n_1)|n_1, n_2\rangle.$$

Wave function corresponding to this coherent state vector has the following Gaussian form

$$\psi(x,y) = \sqrt{\frac{1}{2\pi}} e^{-\frac{(x^2+y^2)}{4}} e^{-\frac{1}{2}(|\alpha|^2+|\beta|^2)+\frac{1}{\sqrt{2}}[\beta(x+iy)+i\alpha(x-iy)]-i\alpha\beta} . \qquad (20)$$

Symplectic tomogram for the coherent state can be given by direct calculations from (8):

$$w^{const} = \frac{e^{-|\alpha|^2-|\beta|^2}}{2\pi\sqrt{\frac{v_1^2}{4}+\mu_1^2}\sqrt{\frac{v_2^2}{4}+\mu_2^2}} \left| \exp\left( \frac{\left(\frac{\beta+i\alpha}{\sqrt{2}} - \frac{iX_1}{v_1}\right)^2}{1-\frac{2i\mu_1}{v_1}} + \frac{\left(\frac{i\beta+\alpha}{\sqrt{2}} - \frac{iX_2}{v_2}\right)^2}{1-\frac{2i\mu_2}{v_2}} - i\alpha\beta \right) \right|^2$$

(21)

This tomogram can be also presented in the form of normal distribution of two random variables:

$$w^{const} = \frac{1}{2\pi\sqrt{\left(v_1^2+\frac{\mu_1^2}{4}\right)\left(v_2^2+\frac{\mu_2^2}{4}\right)}} \exp\left( -\frac{1}{2}(\tilde{X}_1, \tilde{X}_2) \begin{pmatrix} \frac{1}{v_1^2+\frac{\mu_1^2}{4}} & 0 \\ 0 & \frac{1}{v_2^2+\frac{\mu_2^2}{4}} \end{pmatrix} \begin{pmatrix} \tilde{X}_1 \\ \tilde{X}_2 \end{pmatrix} \right),$$

(22)

where $\tilde{X}_i = X_i - \langle X_i \rangle_{\bar{\mu},\bar{v}}$ and $\langle X_1 \rangle = \text{Re}\frac{(v_1+2i\mu_1)(\alpha-i\beta)}{\sqrt{2}}$, $\langle X_2 \rangle = \text{Re}\frac{(v_2+2i\mu_2)(\beta-i\alpha)}{\sqrt{2}}$.

*Coherent states of charged particle moving in uniform varying electromagnetic field*

In this section we consider the charge moving in a constant magnetic field $H$, directed along the Z-axis and varying electric field with the components

$E_x = E_1(t)$ and $E_y = E_2(t)$, $E_z = 0$.

In fact, when the electromagnetic field is varying in time the time-dependence of electric field $\vec{E}(t)$ induces the magnetic field due to Maxwell equations. In view of this the problem which we consider has to be formulated more accurately in the sense that for the Hamiltonian describing the system we choose vector potential and scalar potential in such a form that Schrodinger evolution equation for charged particle wave function $\psi(x,y,t)$ has the form:

$$i\frac{\partial \psi}{\partial t} = -\frac{1}{2}\left(\frac{\partial^2 \psi}{\partial x^2}+\frac{\partial^2 \psi}{\partial y^2}\right)+\frac{1}{8}\left(x^2+y^2\right)\psi +\frac{i}{2}\left(x\frac{\partial \psi}{\partial y}-y\frac{\partial \psi}{\partial x}\right)-\left(E_1 x+E_2 y\right)\psi . \qquad (23)$$

Thus we assume that external sources of charge and current densities provide the field under consideration.

Then one can find the integrals of motion which are called dynamical invariants (see [3], [41], [43]). The invariants will be represented as follows [3]

$$\hat{A}^{var} = \frac{i\exp\left(\frac{it}{2}\right)}{\sqrt{2}}\left[\sqrt{\frac{1}{2}}(z+z_0)+\sqrt{2}\left(\frac{\partial}{\partial \bar{z}}+i\dot{z}_0\right)\right],$$

$$\hat{B}^{var} = -\frac{\exp\left(\frac{it}{2}\right)}{\sqrt{2}}\left[\sqrt{\frac{1}{2}}(\bar{z}+\bar{z}_0)+\sqrt{2}\left(\frac{\partial}{\partial z}+i\dot{\bar{z}}_0\right)\right], \qquad (24)$$

where $z = -\frac{x+iy}{\sqrt{2}}\exp\left(\frac{it}{2}\right)$, $\bar{z} = -\frac{x-iy}{\sqrt{2}}\exp\left(-\frac{it}{2}\right)$.

So the invariants are operators given in Schrodinger representation and containing the time-dependence in explicit form.

The complex quantity $z_0$ depends on time and corresponds the solution of classical motion equation of charge in the electromagnetic field under consideration

$$\ddot{z}_0 +\frac{1}{4}z_0 = F, \ F = \frac{1}{\sqrt{2}}\left(E_1+iE_2\right)e^{\frac{it}{2}}.$$

The dynamical invariants (24) provide the integrals of motion $\hat{A}^+\hat{A}$ and $L_z = \hat{B}^+\hat{B}-\hat{A}^+\hat{A}$ which are generalization of invariants of the charge moving in constant magnetic field. For the charge moving in presence of electric field the invariants contain time-dependence. The physical meaning of these invariants is the following: the invariant $\hat{A}^+\hat{A}$ determines the initial energy of the charge and the invariant $L_z = \hat{B}^+\hat{B}-\hat{A}^+\hat{A}$ determines the initial angular momentum.

We can construct the coherent state corresponding to the Schrödinger equation (23)

$$\langle x,y|\alpha,\beta\rangle = \sqrt{\frac{1}{2\pi}}\exp\left[-\frac{1}{2}\left(|z|^2+|z_0|^2\right)-\left(i\dot{\bar{z}}_0+\frac{1}{2}\bar{z}_0\right)z-\left(i\dot{z}_0+\frac{1}{2}z_0\right)\bar{z}-\frac{it}{2}-\frac{|\alpha|^2}{2}-\frac{|\beta|^2}{2}-i\alpha\beta e^{-it}-\right.$$

$$\left. -ie^{\frac{it}{2}}\left[\alpha(\bar{z}+\bar{z}_0)-i\beta(z+z_0)\right]+i\int_0^t\left(\frac{1}{4}|z_0|^2-|\dot{z}_0|^2\right)d\tau\right]. \qquad (25)$$

The coherent states are eigenstates of the integrals of motion introduced above with eigenvalues $\alpha$ and $\beta$ respectively. Now we can calculate the tomographic probability distribution of the charge coherent state.

Thus, the symplectic tomogram is obtained by direct calculations of the integral (8) and we get a normal distribution of two variables.

$$w^{\text{var}} = \frac{e^{-|\alpha|^2 - |\beta|^2 - |z_0|^2}}{2\pi\sqrt{\left(\frac{1}{4}\nu_1^2 + \mu_1^2\right)\left(\frac{1}{4}\nu_2^2 + \mu_2^2\right)}} \times$$

$$\times \left|\exp\left[\frac{1}{2}\frac{\left(c_1 e^{\frac{it}{2}} + c_2 e^{-\frac{it}{2}} - \frac{iX_1}{\nu_1}\right)^2}{\frac{1}{2} - \frac{i\mu_1}{\nu_1}} + \frac{1}{2}\frac{\left(ic_1 e^{\frac{it}{2}} - ic_2 e^{-\frac{it}{2}} - \frac{iX_2}{\nu_2}\right)^2}{\frac{1}{2} - \frac{i\mu_2}{\nu_2}}\right]\right|^2 \left|e^{-\frac{1}{2}\Lambda D \Lambda^T + \Lambda \vec{l}}\right|^2 \quad (26)$$

where we introduced the two-vector $\vec{\Lambda} = (\alpha \quad \beta)$, the $2 \times 2$-matrix

$$D = \begin{pmatrix} \frac{e^{-2it}b}{2} & ie^{-it}\left(1 - \frac{a}{2}\right) \\ ie^{-it}\left(1 - \frac{a}{2}\right) & -\frac{1}{2}b \end{pmatrix}$$ and the two-vector

$$\vec{l} = \frac{1}{\left(1 - \frac{a}{2}\right)^2 - \left(\frac{b}{2}\right)^2} \begin{pmatrix} -\frac{ie^{it}}{\sqrt{2}}\left(1 - \frac{a}{2} + \frac{b}{2}\right)\tilde{c}_1 + \frac{e^{it}}{\sqrt{2}}\left(1 - \frac{a}{2} - \frac{b}{2}\right)\tilde{c}_2 + ie^{\frac{it}{2}}\left(\overline{z}_0 e^{it}\frac{b}{2} + z_0\left(1 - \frac{a}{2}\right)\right) \\ \frac{1}{\sqrt{2}}\left(1 - \frac{a}{2} + \frac{b}{2}\right)\tilde{c}_1 - \frac{i}{\sqrt{2}}\left(1 - \frac{a}{2} - \frac{b}{2}\right)\tilde{c}_2 - e^{\frac{it}{2}}\left(\overline{z}_0\left(1 - \frac{a}{2}\right) + z_0 e^{-it}\frac{b}{2}\right) \end{pmatrix},$$

where the parameters $a$, $b$ and the functions of time $c_1$, $c_2$, $\tilde{c}_1$ and $\tilde{c}_2$ are given by relations:

$$a = \frac{1}{\frac{1}{2} - \frac{i\mu_1}{\nu_1}} + \frac{1}{\frac{1}{2} - \frac{i\mu_2}{\nu_2}}, \quad b = \frac{1}{\frac{1}{2} - \frac{i\mu_1}{\nu_1}} - \frac{1}{\frac{1}{2} - \frac{i\mu_2}{\nu_2}},$$

$$\tilde{c}_1 = \frac{c_1 e^{\frac{it}{2}} + c_2 e^{-\frac{it}{2}} - \frac{iX_1}{\nu_1}}{\frac{1}{2} - \frac{i\mu_1}{\nu_1}}, \quad \tilde{c}_2 = \frac{ic_1 e^{\frac{it}{2}} - ic_2 e^{-\frac{it}{2}} - \frac{iX_2}{\nu_2}}{\frac{1}{2} - \frac{i\mu_2}{\nu_2}},$$

$$c_1 = \frac{i\dot{\overline{z}}_0 + \frac{1}{2}\overline{z}_0}{\sqrt{2}}, \quad c_2 = \frac{i\dot{z}_0 + \frac{1}{2}z_0}{\sqrt{2}}.$$

Representing the symplectic tomogram in the Gaussian form analogously to the sec. 3 we yield the same formula

$$w^{var} = \frac{1}{2\pi\sqrt{\left(v_1^2 + \frac{\mu_1^2}{4}\right)\left(v_2^2 + \frac{\mu_2^2}{4}\right)}} \exp\left(-\frac{1}{2}(\tilde{X}_1, \tilde{X}_2)\begin{pmatrix} \frac{1}{v_1^2 + \frac{\mu_1^2}{4}} & 0 \\ 0 & \frac{1}{v_2^2 + \frac{\mu_2^2}{4}} \end{pmatrix}\begin{pmatrix} \tilde{X}_1 \\ \tilde{X}_2 \end{pmatrix}\right)$$

with the average values of the variables:

$$\langle X_1 \rangle = \mathrm{Re}\left(\frac{(e^{-it}\alpha - i\beta)}{\sqrt{2}} - i\left(c_1 e^{\frac{it}{2}} + c_2 e^{-\frac{it}{2}}\right)\right)(v_1 + 2i\mu_1), \qquad (27)$$

$$\langle X_2 \rangle = \mathrm{Re}\left(\frac{(-ie^{-it}\alpha + \beta)}{\sqrt{2}} + \left(c_1 e^{\frac{it}{2}} - c_2 e^{-\frac{it}{2}}\right)\right)(v_2 + 2i\mu_2) \qquad (28)$$

Using the property of coherent state to be generating function for Fock state (see [41]) we can calculate the tomogram of the Fock state considering (26). Thus the tomogram of a state with discrete observables $n_1$ and $n_2$ will be determined by Hermite polinimial of two variables [41]

$$w_{n_1 n_2}^{var} = \frac{1}{n_1! n_2!} \frac{e^{-|z_0|^2}}{2\pi\sqrt{\left(\frac{1}{4}v_1^2 + \mu_1^2\right)\left(\frac{1}{4}v_2^2 + \mu_2^2\right)}} \times$$

$$\times \left|\exp\left[\frac{1}{2}\frac{\left(c_1 e^{\frac{it}{2}} + c_2 e^{-\frac{it}{2}} - \frac{iX_1}{v_1}\right)^2}{\frac{1}{2} - \frac{i\mu_1}{v_1}} + \frac{1}{2}\frac{\left(ic_1 e^{\frac{it}{2}} - ic_2 e^{-\frac{it}{2}} - \frac{iX_2}{v_2}\right)^2}{\frac{1}{2} - \frac{i\mu_2}{v_2}}\right]\right|^2 \left|H_{n_1 n_2}^{\{D\}}(\vec{l})\right|^2 \qquad (29)$$

## *Conclusion*

To resume we point out the main results of this work. We constructed tomographic probability distributions for maximally classical states (coherent states) of quantum charged particle moving in uniform time-varying electric field and constant magnetic field. The probability distribution is shown to be normal distribution with time-dependent parameters. We also found the tomographic probability distributions describing the states with discrete quantum numbers which are eigenstates of time-dependent integrals of motion, namely, the initial energy and initial angular momentum of the charge moving in the time-varying electric field and constant magnetic field.

In next work we will consider the transition probabilities between the Landau levels induced by time-varying field in the tomographic probability representation.

*Acknowledgements*

The authors thank the Russian Foundation for Basic Research for partial support under Project Nos. 10-02-00312.